\documentclass[preprint,12pt]{elsarticle}

\usepackage{amssymb}
\usepackage{epsfig}        
\usepackage{graphicx}	
\usepackage{xcolor}
\usepackage{hyperref}

\newcommand{\beq}{\begin{equation}}
\newcommand{\eeq}{\end{equation}}
\newcommand{\beqa}{\begin{eqnarray}}
\newcommand{\eeqa}{\end{eqnarray}}
\newcommand{\bd}[1]{ \mbox{\boldmath $#1$}}
\newcommand{\dd}{\mathrm{d}}

\journal{Nuclear Physics A}

\begin{document}
\def\ii{\'\i}

\begin{frontmatter}

\title{Quantum Phase Transitions within a nuclear cluster
model and an effective model of QCD}

\author{D.S. Lohr-Robles$^{1}$, E. L\' opez-Moreno$^{2}$ and P.O. Hess$^{1,3}$}

\address{
{\small\it
$^1$ Instituto de Ciencias Nucleares, Universidad Nacional Aut\'onoma de M\'exico,}\\
{\small\it A.P. 70-543, 04510 Mexico-City, Mexico}\\
{\small\it $^2$ Facultad de Ciencias,  Universidad Nacional Aut\'onoma de M\'exico,} \\
{\small\it 04510 Mexico-City, Mexico} \\
{\small\it $^3$ Frankfurt Institute for Advanced Studies, J. W. von Goethe University, Hessen, Germany}
}

\begin{abstract}
The catastrophe theory is applied to a nuclear cluster 
model and an effective model for QCD at low 
energy. 
The study of quantum phase transitions in the cluster model was considered in an earlier publication, but restricted to spherical clusters and on a semi-classical level. In the present contribution, we include the case of deformed clusters and determine the spectrum numerically as a function of an interaction parameter, where signatures of a quantum phase transition can be seen.
It is shown that in this more complicated case, with
deformation of the clusters, the
catastrophe theory can be applied with some
interesting consequences.
A further example of a many-body problem
is considered, namely
an effective model of QCD, which is able to describe
the low energy hadron spectrum and, when temperature is 
introduced, even ratios of particle-antiparticle 
productions. The catastrophe theory is able
to provide useful information on the phase
transition from a perturbative to a non-perturbative
vacuum. This contributions shows the universal
usefulness of catastrophe theory, while more 
examples of applications to different fields are
mentioned in the Introduction.
\end{abstract}


\end{frontmatter}

\section{Introduction}
\label{sec1}

In a recent contribution \cite{NPA2019} we applied the
catastrophe theory \cite{gilmore} to 
the {\it Semimicroscopic Algebraic Cluster Model}
(SACM), in order to describe {\it Quantum Phase Transitions}
(QPTs). The SACM was chosen because it exhibits quite
complex semi-classical potentials and, therefore,
serves as a test-bed for the applied method and how
to proceed in more complex models. 
In \cite{NPA2019} the study of QPTs was limited to spherical clusters on a semi-classical level. In this contribution we include deformed clusters, which introduces a higher complexity in the semi-classical analysis. In addition, explicit numerical diagonalisations of the Hamiltonian are done and the spectrum is retrieved as function of an interaction parameter, which will show characteristics related to the phase transitions.

Another motivation of this contribution is to 
show the effectiveness of the catastrophe theory
and its wide range of applications to quite different fields
in physics, all related to many-particle physics
but from a different perspective.
We will apply it to an effective model of QCD at low 
energy. First steps into this direction were presented
in \cite{QCD-0,QCD-1,QCD-2,QCD-3}, using standard methods.
With the catastrophe theory we can deduce the order of the
phase transition and describe important changes in the
structure of the vacuum  state of QCD.

There are applications even to General Relativity.
For example, in \cite{universum} the catastrophe theory
was applied to a rotating black hole, with a Kerr metric.
Phase transitions were related to the appearing and
disappearing  of the event-horizon of black holes and to the positions of 
light rings. A further application can be found in
Optics and we refer to the book \cite{optics}.

The paper is structured as follows: 
In Section \ref{sec2} we discuss the extension of the SACM to include deformation of the clusters and the obtention of the semi-classical potential. Two examples are considered, one of spherical clusters and one of deformed clusters, and the study of QPTs, semi-classically and numerically, is done.
In Section \ref{QCD} the
catastrophe theory will be applied to an effective model of 
QCD at low energy and the many-body structure of the
vacuum state is investigated.
In Section \ref{conclusions}
Conclusions will be drawn.

\section{QPTs within the SACM and their signatures}
\label{sec2}

Cluster models play an important role in nuclear physics.
There are two main groups of cluster models: The
microscopic cluster models, which can be represented by 
\cite{schuck-CM}, and algebraic cluster models. 
The last group
can be divided into the ones which do not satisfy the
{\it Pauli Exclusion Principle} (PEP), for example
\cite{bijker-ann}, and those which do observe the PEP,
for example the SACM \cite{sacm1,sacm2}. 
The importance of satisfying the PEP and the consequences of the PEP not being observed can be found in \cite{12C,16O}.
There is another important line of models able to
study the clusterization of nuclei, namely
\cite{Lau2016,dreyfuss2020,Dytrich2020}, which uses
a {\it Symmetry Adapted} basis in reducing the
shell model space. The cluster structure is investigated
via overlaps of a symplectic basis with cluster wave 
functions.

Phase transitions in algebraic models were discussed 
in  \cite{lopez2}, for the IBA \cite{IBA}, 
and in various other algebraic 
models in \cite{cejnar2007,caprio2008,casten2010,cejnar2010}.
In particular, in \cite{leviathan} the catastrophe theory
was applied to these algebraic models.
Here, we will apply it to the SACM \cite{sacm1,sacm2}.
For completeness, a short summary is presented 
on the algebraic cluster model considered.

The SACM was proposed in 1994 \cite{sacm1, sacm2} and 
differs from most algebraic 
clusters models because it takes into account the PEP. 
The relative motion of the clusters is 
described by the generators of the $\mathrm{U}_R(4)$ group, 
which are the $16$ combinations of the creation 
$({\bd \pi}^{\dagger}_m, {\bd \sigma}^{\dagger})$ and 
annihilation  $({\bd \pi}^m, {\bd \sigma})$ operators of the 
$\pi$ bosons (with angular momentum $\ell =1$) and the 
$\sigma$ bosons (with angular momentum $\ell =0$): 
${\bd \pi}^{\dagger}_m{\bd \pi}^{m'}$, ${\bd \pi}^{\dagger}_m{\bd \sigma}$, ${\bd \sigma}^{\dagger}{\bd \pi}^m$, 
${\bd \sigma}^{\dagger}{\bd \sigma}$.

The Pauli exclusion principle is considered in the way 
on how the space of the SACM is constructed. Each cluster is described by an $(\lambda_k, \mu_k)$ 
irreducible representation (irrep) within the 
$\mathrm{SU}(3)$ shell model, while the relative motion is also described as an $(n_{\pi},0)$ irrep, with the PEP 
being accounted for partially through the Wildermuth condition \cite{wildermuth}, which imposes a minimum value to the number of $\pi$ bosons: $n_{\pi}\geq n_0 $. Then, the space is constructed as the direct product of these irreps,
\begin{equation}\label{1.1}
(\lambda_1, \mu_1)\otimes (\lambda_2, \mu_2)\otimes (n_{\pi},0) = \sum_{(\lambda, \mu)}m_{\lambda, \mu} (\lambda, \mu),
\end{equation}
where $m_{\lambda, \mu}$ is a multiplicity factor. The 
resulting sum of irreps is finally compared with the space of 
the $\mathrm{SU}(3)$ shell model of the nucleus considered
and only those irreps are maintained which appear
in the shell model. In such a manner, the PEP is
accounted for.

The \textit{semi} prefix in the name of the model is because the Hamiltonian operator is phenomenological and is composed of Casimir operators of the dynamical symmetries. 
For the relative motion, 
there are two group chains of $\mathrm{U}(4)$ which contains 
the group $\mathrm{SO}(3)$ of angular momentum:
\begin{eqnarray}\label{1.2}
&&\mathrm{U}(4) \supset \mathrm{SU}(3)\supset \mathrm{SO}(3)\supset \mathrm{SO}(2)\cr
&&\mathrm{U}(4) \supset \mathrm{SO}(4)\supset \mathrm{SO}(3)\supset \mathrm{SO}(2) ~~~,
\end{eqnarray}
defining the dynamical symmetry (here we omitted the $R$  subscript of the relative motion). As an example we consider
a simplified (not the most general)
Hamiltonian operator, consisting of linear combinations of 
Casimir operators of the groups $\mathrm{SU}(3)$ and $\mathrm{SO}(4)$ up to second order:
\begin{equation}\label{1.3}
{\bd H} = {\bd H}_{\mathrm{SU}(3)} + {\bd H}_{\mathrm{SO}(4)}
\end{equation}
with
\begin{eqnarray}
{\bd H}_{\mathrm{SU}(3)} &=& \hbar \omega {\bd n}_{\pi} + (\bar{a}-\bar{b}\Delta{\bd n}_{\pi}){\bd C}_2({\bd n}_{\pi},0) + (a-b\Delta{\bd n}_{\pi}){\bd C}_2(\lambda,\mu)+\xi {\bd L}^2 + t_1 {\bd K}^2  \label{1.3a}\\
{\bd H}_{\mathrm{SO}(4)} &=& \frac{c}{4} \left[({\bd \pi}^{\dagger} \cdot {\bd \pi}^{\dagger}) -({\bd \sigma}^{\dagger})^2 \right] \left[({\bd \pi} \cdot {\bd \pi}) -({\bd \sigma})^2 \right] , \label{1.3b} 
\end{eqnarray} 
with $\Delta{\bd n}_{\pi}= {\bd n}_{\pi} - n_0$, and $n_0$ is the minimal number of relative oscillation quanta
determined by the Wildermuth condition \cite{wildermuth}.
The $\hbar\omega$ is given by $45A^{-\frac{1}{3}}
-25A^{-\frac{2}{3}}$ \cite{hw} and, thus, is fixed. 
The Hamiltonian depends on $7$ parameters 
$\{a,b,\bar{a},\bar{b},c,\xi , t_1\}$, all in units of
MeV. 
The operator ${\bd K}^2$ is introduced to address the degeneracy of excited states.
The eigenvalues of the second order Casimir, the angular
momentum and ${\bd K}^2$ operators are
\beqa
{\bd C}_2(\lambda , \mu ) & \rightarrow &
\lambda^2 + \lambda\mu + \mu^2 + 3\lambda + 3\mu
\nonumber \\
{\bd L}^2 & \rightarrow & L(L+1)
\nonumber \\
{\bd K}^2 & \rightarrow & K^2
~~~,
\label{eigenvalues}
\eeqa
where $K$ is the quantum number labelling the rotational
bands.

After having set up the model Hamiltonian, the next step
is to define the semi-classical potential. As this was already done in \cite{NPA2019}, 
we restrict to a short summary, adding  the contributions due to the deformation
of the clusters.

The semi-classical potential is obtained as the expectation 
value of the Hamiltonian operator in the basis of the 
coherent states of the SACM, which are defined as \cite{geom}
\begin{equation}\label{1.4}
|\alpha \rangle = \frac{N!}{(N+n_0)!}\mathcal{N}_{N,n_0}\left.\frac{d^{n_0}}{d \gamma^{n_0}}[{\bd \sigma}^{\dagger}+\gamma ({\bd \alpha}^{*}\cdot{\bd \pi}^{\dagger})]|0\rangle \right|_{\gamma=1}
\end{equation}
where
\begin{equation}\label{1.5}
\mathcal{N}^{-2}_{N,n_0}=\left. \frac{(N!)^2}{(N+n_0)!}\frac{d^{n_0}}{d \gamma^{n_0}_1}\frac{d^{n_0}}{d \gamma^{n_0}_2}[1+\gamma_1\gamma_2({\bd \alpha}^{*}\cdot{\bd \alpha})]^{N+n_0}\right|_{\gamma_1=\gamma_2=1}
~~~.
\end{equation}
$\mathcal{N}^{-2}_{N,n_0}$
is the normalization constant and $\alpha_m$ are arbitrarily complex variables. 
We consider the simple case in which $\alpha_m$ transforms 
as a tensor, as was previously done in \cite{fraser2012a}, and choose the parametrization $\alpha_{+1}=\alpha_{-1}=0$ and $\alpha_0=i\alpha$, where the variable $\alpha$ can be related to the distance between the clusters \cite{geom}.

The semi-classical potential results in a function of one variable $\alpha$:
\begin{eqnarray}\label{1.7}
 V(\alpha;c_i)&=&\langle \alpha | {\bd H} | \alpha \rangle \cr
 &=& V_0-(b+{\bar b})
\left( A\alpha^2  \frac{F_{11}(\alpha)}{F_{00}(\alpha)}+B\alpha^4  \frac{F_{22}(\alpha)}{F_{00}(\alpha)}+\alpha^6  \frac{F_{33}(\alpha)}{F_{00}(\alpha)}+ C \alpha^2  \frac{F_{20}(\alpha)}{F_{00}(\alpha)} \right),
\end{eqnarray}
with the constant value $V_0$ given by
\begin{equation}\label{1.8}
V_0 = (a+n_0 b)\langle {\bd C}_2(\lambda_C,\mu_C) \rangle + \xi \langle {\bd L}^2_C \rangle + \frac{c}{4}(N+n_0)(N+n_0-1) ,
\end{equation}
where $(\lambda_C,\mu_C)$ is an intermediate irrep of the cluster system given by the product $(\lambda_1, \mu_1)\otimes (\lambda_2, \mu_2)$ in (\ref{1.1}).

The $3$ control parameters $c_i=\{A,B,C\}$ are linear combinations of the Hamiltonian parameters:
\begin{eqnarray}
A&=&-\frac{1}{b+\bar{b}}\left[\hbar \omega - b\langle {\bd C}_2(\lambda_C,\mu_C)\rangle + 4\big(a+\bar{a}+(b+\bar{b})(n_0-1)\big)+ \big(a+b(n_0-1)\big)\left(\Gamma_1+\Gamma_2\right)\right.\cr
&&\left. +2\xi -\frac{c}{2}(N+n_0-1)  \right] \label{1.8a}\cr
B&=&-\frac{1}{b+\bar{b}}\left[ a +\bar{a} + (b+\bar{b})(n_0-6)-b (\Gamma_1+\Gamma_2)+\frac{c}{2}\right]\label{1.8b}\\
C&=&\frac{1}{b+\bar{b}} \frac{c}{2}\label{1.8c}
~~~. \nonumber
\end{eqnarray}
The deformation of the clusters is taken into account via 
the $\Gamma_k$, 
which is the expectation value of the $m=0$ component of the quadrupole operator of cluster $k$ 
\cite{fraser2012a}:
\begin{equation}\label{1.9}
\Gamma_k =\langle (\lambda_k,\mu_k) | Q^{C_k}_0| (\lambda_k,\mu_k)\rangle = \sqrt{\frac{5}{\pi}} \left(n_k + \frac{3}{2}(A_k -1)\right) \beta_k
~~~.
\end{equation}
The $n_k$ is the total number of quanta of the deformed cluster and $\beta_k$ is the quadrupole deformation. The $F_{pq}(\alpha^2)$ functions in (\ref{1.7}) are defined in \cite{fraser2012a}:
\begin{eqnarray}\label{1.10}
F_{pq}(\alpha^2)&=& \frac{(N!)^2}{(N+n_0-\max(p,q))!}\\
&&\times\sum_{k=\max(n_0-p,n_0-q)}^{N+n_0-\max(p,q)}\left(
\begin{array}{c}
N+n_0-\max(p,q) \\
k
\end{array}
\right)
\frac{(k+p)!}{(k+p-n_0)!}\frac{(k+q)!}{(k+q-n_0)!}\alpha^{2k} .\nonumber
\end{eqnarray}

\subsection{Catastrophe theory and the parameter space}

A useful and systematic way of determining how the change of parameters affects a function's critical points is with the help of catastrophe theory \cite{gilmore}. 
For details, concerning the first application to
the SACM, please consult \cite{NPA2019}. 

The parameter space will be divided into regions of similar qualitative behaviour of the potential. Two separatrices,
which divide regions of different behaviours, of importance are to be constructed: The bifurcation and the Maxwell sets. 
\begin{itemize}
\item Bifurcation set: Is the subspace of parameter space delimiting the emergence of extreme values of the potential.
\item Maxwell set: Is the subspace of parameter space where two or more extreme values of the potential are the same, i.e. for $\alpha_1$ and $\alpha_2$ critical points we have $V(\alpha_1)=V(\alpha_2)$.
\end{itemize}
In \ref{appendix1} a general expressions for the calculations of the bifurcation set and the Maxwell set, for an arbitrary potential function, are presented.

\begin{figure}[ht]
\begin{center}
\includegraphics[scale=1]{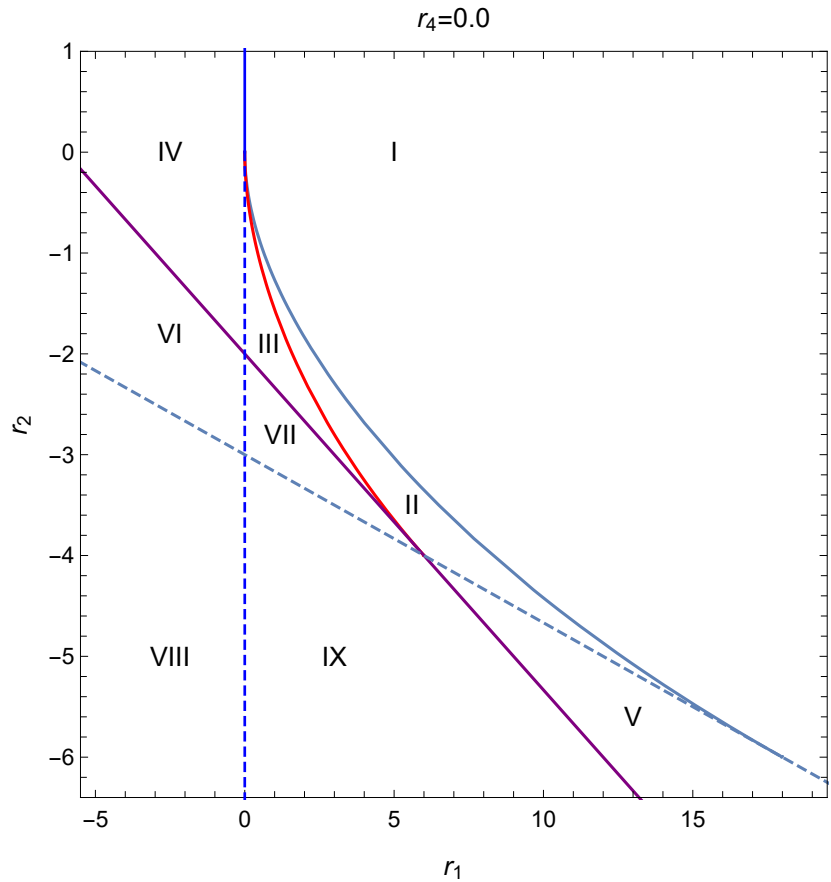}
\end{center}
\caption{Parameter space $(r_1,r_2)$. The $r_2$-axis is drawn as a blue continuous line for $r_2>0$, where $\alpha =0$ is a minimum point, and as a blue dashed line for $r_2<0$, where $\alpha=0$ is a maximum point; in both cases the critical point has a fourth order multiplicity. The bifurcation set is drawn as a light blue line, and the Maxwell set as a red line. The light blue dashed line is the continuation of the bifurcation set, which is obtained by joining the end of the bifurcation set to the end of the Maxwell set. The stability separatrix in (\ref{2.8}) is drawn as a purple line. The particular characteristics which define each region are described in the text.}
\label{figure1}
\end{figure}

A first step in the catastrophe theory formalism is the determination of the essential parameters \cite{lopez2}, which are the minimum number of parameters necessary for a complete description
of the potential and are combinations of the original control parameters. They are obtained by expanding the potential in a Taylor series about the fundamental root. We may identify the origin $\alpha =0$ as the fundamental root since it is always a critical point of the semi-classical potential (\ref{1.7}) for all values of the control parameters. Then, by choosing combinations of the control parameters such that the first terms in the Taylor series vanish, until it is no longer possible to eliminate the next term, which is called the \textit{germ} of the potential, we determine the essential parameters.

A Taylor series expansion of the semi-classical potential (\ref{1.7}) about $\alpha =0$ yields:
\begin{equation}\label{2.1}
V(\alpha;c_i)= V_0 -(b+\bar{b}) (T_0+T_1\alpha^2+T_2\alpha^4+\ldots),
\end{equation}
and the first $T_i$ coefficients are given by
\begin{eqnarray}\label{2.2}
T_0 &=& n_0 A + n_0(n_0-1)B+n_0(n_0-1)(n_0-2)\cr
T_1 &=& N(n_0+1) A + 2N (n_0+1)n_0B + 3N (n_0+1)n_0(n_0-1)\cr
&&+\frac{1}{2}N(N-1)(n_0+2)(n_0+1)C \cr
T_2 &=& N(n_0+1)\Big(N-(n_0+2)\Big)A + N(n_0+1)\Big(2(N-1)-n_0(2n_0+5-3N)\Big)B\cr
&&+3N(n_0+1)n_0\Big(N(2n_0+1)-2n_0(n_0+1)\Big)\cr
&&+\frac{1}{3}N(N-1)(n_0+2)(n_0+1)\Big(n_0(N+1)+3\Big)C .
\end{eqnarray}
The semi-classical potential is then redefined as
\begin{equation}\label{2.3}
\bar{V}(\alpha;c_i)=V(\alpha;c_i)-V_0+(b+\bar{b})T_0,
\end{equation}
so that we have $\bar{V}(\alpha\to 0;c_i)=0$. By straightforward algebraic manipulation of the $F_{pq}(\alpha)$ functions we are able to write the semi-classical potential as:
\begin{equation}\label{2.4}
\bar{V}(\alpha, r_i) =  \frac{r_3}{Q_0(\alpha)}\left[r_1 Q_1(\alpha) + r_2 Q_2 (\alpha) + Q_3(\alpha)+ r_4 Q_4(\alpha) \right],
\end{equation}
with the essential parameters $r_i$ defined by
\begin{eqnarray}\label{2.5}
r_1 &=& A+2n_0 B +3n_0(n_0-1)+\frac{1}{2}(N-1)(n_0+2)C\cr
r_2 &=& B+3n_0 -\frac{1}{6}(n_0(N+1)+6)C \cr
r_3 &=& -(b+\bar{b})\cr
r_4 &=& \frac{1}{6}n_0(N+1)C ,
\end{eqnarray}
and define the $Q_i(\alpha)$ polynomials by
\begin{eqnarray}\label{2.6}
Q_i(\alpha) &=& \sum_{k=i}^{N} \frac{N!}{(N-k)!}\frac{(n_0+k)!}{n_0!} \frac{\alpha^{2k}}{k!(k-i)!} \cr
Q_4(\alpha) &=& \sum_{k=3}^{N} \frac{N!}{(N-k)!}\frac{(n_0+k)!}{n_0!} \frac{\alpha^{2k}}{(k+1)!(k-3)!}, 
\end{eqnarray}
with $i=1,2,3$. In (\ref{2.2}) and (\ref{2.5}) we can see that from $T_1=0$ we obtain $r_1$, and similarly from $T_2=0$ we get $r_2$; lastly, $r_3$ and $r_4$ are the leftover parameters. The next term of the Taylor series $T_3$ can no longer be eliminated, thus identifying the germ of the potential as $\alpha^6$.

The stability of the potential is determined 
through the limits $\alpha \to \infty$ and $N\to\infty$. A 
potential is stable if in the limit $\alpha \to\infty$ it 
tends to a positive value, otherwise, if it tends to a 
negative value the potential is unstable
(the system disintegrates when approaching
infinity). Using the 
expressions in (\ref{2.6}) we readily obtain the limit 
$\alpha\to\infty$ of the semi-classical potential (\ref{2.4}):
\begin{equation}\label{2.7}
\lim_{\alpha\to\infty}V(\alpha;r_i)= r_3N\left(r_1 +r_2(N-1)+(N-1)(N-2) + r_4 \frac{(N-1)(N-2)}{N+1} \right),
\end{equation}
and the global $N$ indicates that in the limit $N\to\infty$ the potential will either tend to plus (stable) 
or minus (unstable) infinity. 
By demanding (\ref{2.7}) to be zero we obtain the separatrix 
in parameter space
\begin{equation}\label{2.8}
 r_1 +r_2(N-1)+(N-1)(N-2) + r_4 \frac{(N-1)(N-2)}{N+1}=0,
\end{equation}
which divides the parameter space in regions of stable and unstable potentials.

In the following sections we will study the quantum phase transitions of two separate cases:

\begin{itemize}

\item The limit $\mathrm{SU}(3)$ of the Hamiltonian 
(\ref{1.3}), i.e. $c=0$, when the 
essential parameter space is two-dimensional $(r_1,r_2)$. This part is mostly a repetition of 
\cite{NPA2019}.
However, it contains some new elements: A series of paths for two example systems in the parameter space and their effects as avoided level crossing of states as a function of $\bar{a}$ are considered.

\item The Hamiltonian (\ref{1.3}) has a mixture of $\mathrm{SU}(3)$ and $\mathrm{SO}(4)$ symmetries, i.e. $c\neq 0$ and the essential
parameter space is three-dimensional $(r_1,r_2,r_4)$.
\end{itemize}

\subsection{QPTs in the $\mathrm{SU}(3)$ limit: $c=0$}

For this case, two example systems will be considered: The 
system ${}^{16}\mathrm{O}+\alpha \to {}^{20}\mathrm{Ne}$ of 
spherical clusters, and the system 
${}^{12}\mathrm{C}+{}^{12}\mathrm{C} \to {}^{24}\mathrm{Mg}$ 
of deformed clusters. For each case the separatrices in the 
two-dimensional $(r_1,r_2)$ parameter space are constructed 
and QPTs are studied as one moves across different regions. 
The ${}^{20}\mathrm{Ne}$ clusterisation
was considered in \cite{NPA2019}, while the ${}^{24}\mathrm{Mg}$
system is new, containing 
two well deformed cluster, i.e.,
we can study the effects of deformed clusters, not 
present in the first system. Another new ingredient
is the numerical calculation of the spectrum, relating
its structure to the appearance of phase transitions,
as will be described by the semi-classical analysis.
The software MATHEMATICA \cite{mat11} was extensively used.

\subsubsection{Spherical clusters example: ${}^{16}\mathrm{O}+\alpha \to {}^{20}\mathrm{Ne}$}

For this example we have $n_0=8$, 
deformations $\beta_{{}^{16}\mathrm{O}}=0$, 
$\beta_{\alpha}=0$, 
$(\lambda_1,\mu_1)=(\lambda_2,\mu_2)=(0,0)$ and 
in the numerical diagonalisation of the
Hamiltonian we will consider up to 
four excitation quanta, $N=4$. With the values for $n_0$ and $N$ and direct application of the formulas in \ref{appendix1}, we are able to construct the separatrices in parameter space $(r_1,r_2)$ depicted in Fig. \ref{figure1}. For regions I, II, III, IV and V the potentials tend to a positive value in the limit $\alpha\to\infty$: $V(\alpha\to\infty)>0$; and for regions VI, VII, VIII and IX the potentials tend to a negative value in this limit 
$\alpha\to\infty$: $V(\alpha\to\infty)<0$. In region I there is only one minimum at $\alpha=0$. In region II there are two minima one at $\alpha=0$ and one at $\alpha>0$ with $V(\alpha=0)<V(\alpha>0)$. In region III there are two minima one at $\alpha=0$ and one at $\alpha>0$ with $V(\alpha=0)>V(\alpha>0)$. In region IV there is one maximum at $\alpha=0$ and one minimum at $\alpha>0$. In region V there is 
a minimum at $\alpha =0$, a maximum at $\alpha>0$ and the other minimum is at $\alpha\to\infty$. In region VI there is a maximum at $\alpha=0$ and a minimum at $\alpha>0$. In region VII there is a minimum at $\alpha=0$ and a minimum at $\alpha>0$ with $V(\alpha=0)>V(\alpha>0)$. In region VIII there is only a maximum at $\alpha=0$. In region IX there is a minimum at $\alpha=0$, a maximum at $\alpha>0$ and the other minimum is at $\alpha\to\infty$.

\begin{figure}[ht]
\begin{center}
\includegraphics[scale=0.66]{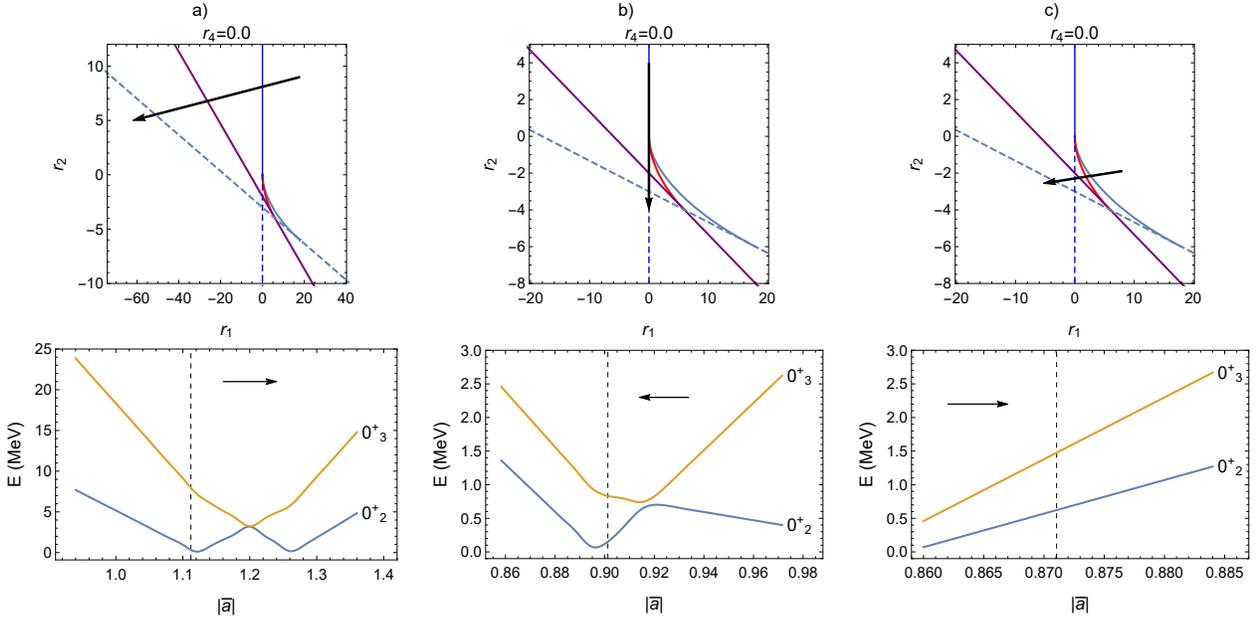}
\end{center}
\caption{Shown are different trajectories in
the $(r_1,r_2)$ parameter space
for ${}^{16}\mathrm{O}+\alpha \to {}^{20}\mathrm{Ne}$. In the top row are the parameter spaces with the respective path of the trajectory depicted as an arrow. In the bottom row we plot the first $0^+$ energy levels as a function of the absolute value of parameter $\bar{a}$, and as an arrow also indicate the direction of the path taken. The vertical dashed line indicates the value of $\bar{a}$ where the trajectory is at the point of a phase transition: In a) crossing from region I to region IV, in b) crossing the point $(r_1=0,r_2=0)$, and in c) crossing from region II to region III. The values of the parameters used are: In a) $\bar{b}=-0.08$; in b) $r_1=0$; and in c) $\bar{b}=-0.036$. For all cases we used: $\xi=0.208$ and $a=b=t_1=0$.}
\label{figure2}
\end{figure}

A second order QPT occurs in a trajectory in parameter space going from region I to region IV. This is because the global minimum of the potential at $\alpha=0$ disappears and becomes a global minimum at $\alpha>0$ as the parameter $r_1$ goes from $r_1>0$ to $r_1<0$ at a fixed $r_2>0$. Following Ehrenfest's classification of phase transitions and 
encountering a discontinuity in the 
second derivative of the global minimum 
with respect to the parameter $r_1$,
we can conclude that the phase transition is
of second order. Similarly, a third order QPT occurs in a trajectory going from $r_2>0$ to $r_2<0$ at $r_1=0$, passing through $r_2=0$, where the potential at $\alpha=0$ goes as $\alpha^6$. Here we encounter a discontinuity in the third derivative of the global minimum of the potential with respect to the parameter $r_2$. Lastly, a first order QPT occurs in a trajectory going from region II to region III, crossing the Maxwell set. Here, the minimum at $\alpha>0$ in region II becomes the global minimum in region III, and we encounter a discontinuity in the first derivative of the global minimum of the potential with respect to the parameters as the transition takes place.

In Fig. \ref{figure2} we show examples of three different trajectories in the $(r_1,r_2)$ parameter space, with their respective plot of the first $0^+$ energy levels as a function of the absolute value of $\bar{a}$. The trajectories in a) and c) are obtained by fixing the parameters $\xi=0.208$, $a=b=t_1=0$, and using $\bar{b}=-0.08$ and $\bar{b}=-0.036$, respectively; the parameter $\bar{a}$ is then varied as shown in the plots. The trajectory in b) is obtained by fixing $r_1=0$ and going from $r_2>0$ to $r_2<0$. For cases a) and b) avoided level crossings in the vicinity of the QPT are found, while in case c) there are no avoided level crossings. This result shows that the signatures for a phase transition depend on the path taken.

\subsubsection{Deformed clusters example: ${}^{12}\mathrm{C}+{}^{12}\mathrm{C} \to {}^{24}\mathrm{Mg}$}

This example allows to include a deformation
dependence on the clusters. 
We have $n_0=12$, $\beta_{{}^{12}\mathrm{C}}=-0.38$, 
$(\lambda_1,\mu_1)=(\lambda_2,\mu_2)=(0,4)$, and 
$(\lambda_C,\mu_C)=(0,8)$.
In the numerical diagonalisation, up to
four excitation quanta $N=4$ are considered.

\begin{figure}[ht]
\begin{center}
\includegraphics[scale=0.8]{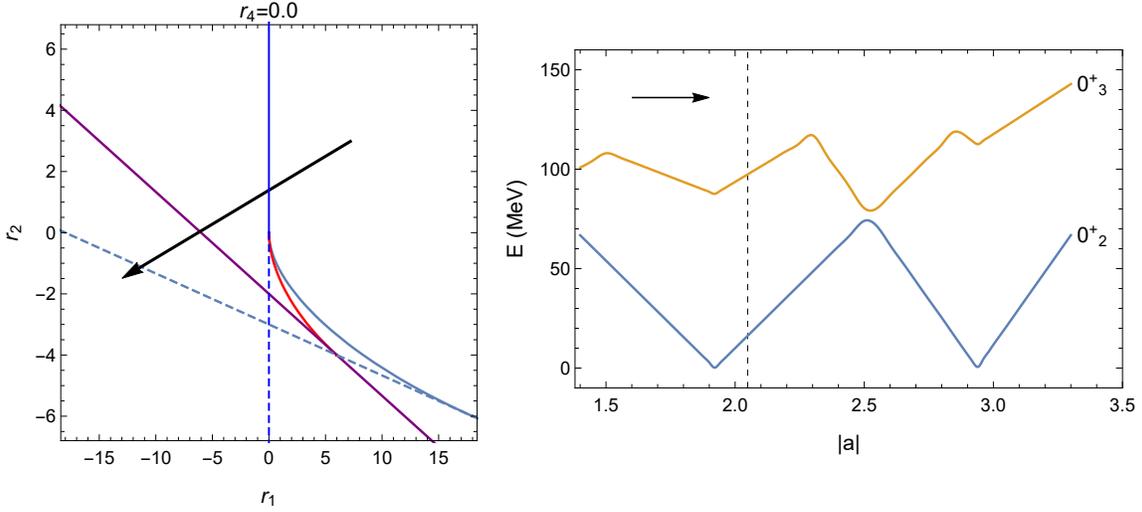}
\end{center}
\caption{In the left figure the trajectory 
is shown within the parameter space $(r_1,r_2)$ 
of a second order QPT for 
${}^{12}\mathrm{C}+{}^{12}\mathrm{C}\to {}^{24}\mathrm{Mg}$ 
in the $\mathrm{SU}(3)$ limit. 
In the right we plot the $0^{+}_2$ and $0^{+}_3$ energy levels 
as a function of the absolute value of the parameter $a$. 
Only the first three $0^+$ states are plotted, including the 
ground state. Note that the
$0_3^+$ state exhibits turns which are the
consequences of further crossings with higher excited
states. The vertical dashed line indicates the value of $a$ where the trajectory crosses from region I to region IV and a phase transition takes place.
The values of the parameters used are: $b=-0.4$, 
$\xi=0.196$, $\bar{a}=\bar{b}=0$, and $t_1=0.7175$.
The arrow indicates the path of the trajectory.}
\label{figure3}
\end{figure}

Similarly to the previous example, a second order QPT occurs 
as the strength of the second order Casimir operator parameter $a$ increases. The left plot in Fig. \ref{figure3} shows the parameter space for the ${}^{12}\mathrm{C}+{}^{12}\mathrm{C}\to {}^{24}\mathrm{Mg}$ system and the particular trajectory taken. The trajectory is obtained by fixing the parameters $b=-0.4$, $\xi=0.196$, $\bar{a}=\bar{b}=0$, 
$t_1=0.7175$, and varying the parameter $a$ from $a=-1.4$ to $a=-3.3$, crossing the $r_2$-axis at approximately $a=-2.0$. In the right plot we depict the $0^{+}_2$ and $0^{+}_3$ energy levels and see that the first avoided energy level crossing of the $0^{+}_2$ with the ground state $0^{+}_1$ occurs at the vicinity of the QPT. The upper turns, seen in the 
curve of $0_3^+$, are due to avoided level crossings with higher lying $0^+$ states, not plotted.

\subsection{QPTs in a Hamiltonian with $\mathrm{SU}(3)$ and $\mathrm{SO}(4)$ symmetry: $c \neq 0$}

\begin{figure}[ht]
\begin{center}
\includegraphics[scale=0.9]{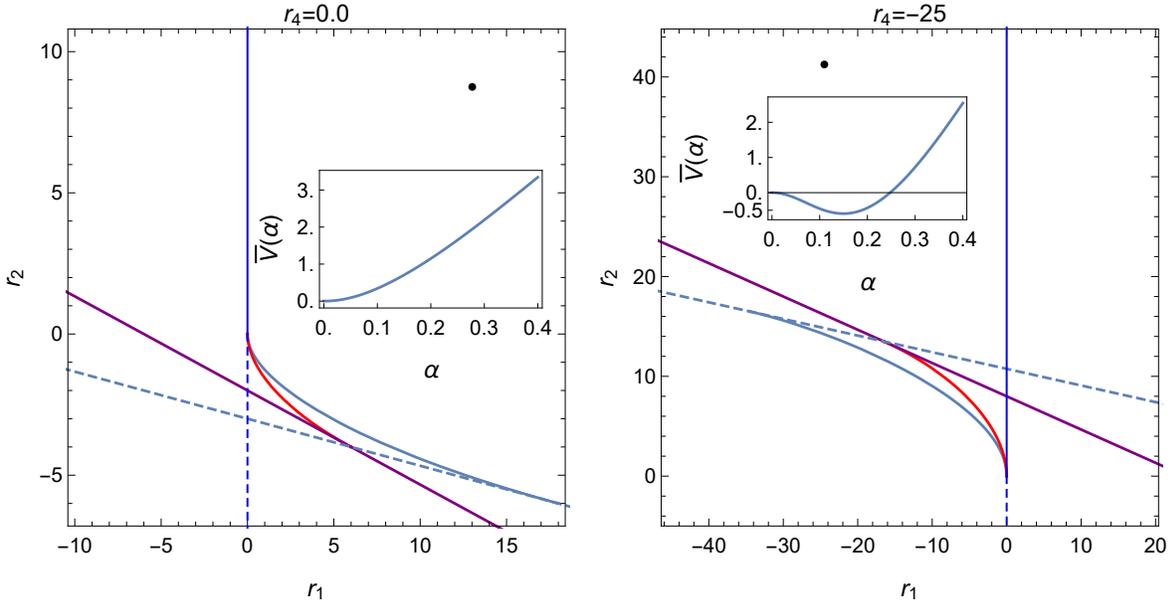}
\end{center}
\caption{
Shown are two slices of parameter space $(r_1,r_2,r_4)$ for 
different values of parameter $r_4$ for the example 
${}^{16}\mathrm{O}+\alpha\to{}^{20}\mathrm{Ne}$. 
The plot in the left corresponds to a point before 
the QPT at $c=0$. The plot in the right corresponds to a 
point after the QPT at $c=0.6$. In both cases an inset with 
the corresponding semi-classical potential is shown. The values of the parameters used are: $\bar{a}=-1.06$, $\bar{b}=-0.08$, $\xi=0.208$, and $a=b=t_1=0$.}
\label{figure4}
\end{figure}

In this case, we focus on the previous example of the system of spherical cluster ${}^{16}\mathrm{O}+\alpha \to {}^{20}\mathrm{Ne}$ and this time turn on the parameter $c$, and search in the three-dimensional parameter space $(r_1,r_2,r_4)$ for a suitable trajectory. Similarly to the $\mathrm{SU}(3)$ limit, we fix the following parameters: $\bar{a}=-1.06$, $\bar{b}=-0.08$, $\xi=0.208$, and $a=b=t_1=0$ and obtain the QPT trajectory varying the parameter $c$, 
which now crosses the $(r_2,r_4)$-plane. In 
Fig. \ref{figure4} we show two slices of parameter space 
$(r_1,r_2,r_4)$ for different values of $r_4$, one at 
$r_4=0$ and the other at $r_4=-25$, which correspond to 
points before and after the second order QPT, respectively.

\begin{figure}[ht]
\begin{center}
\includegraphics[scale=0.7]{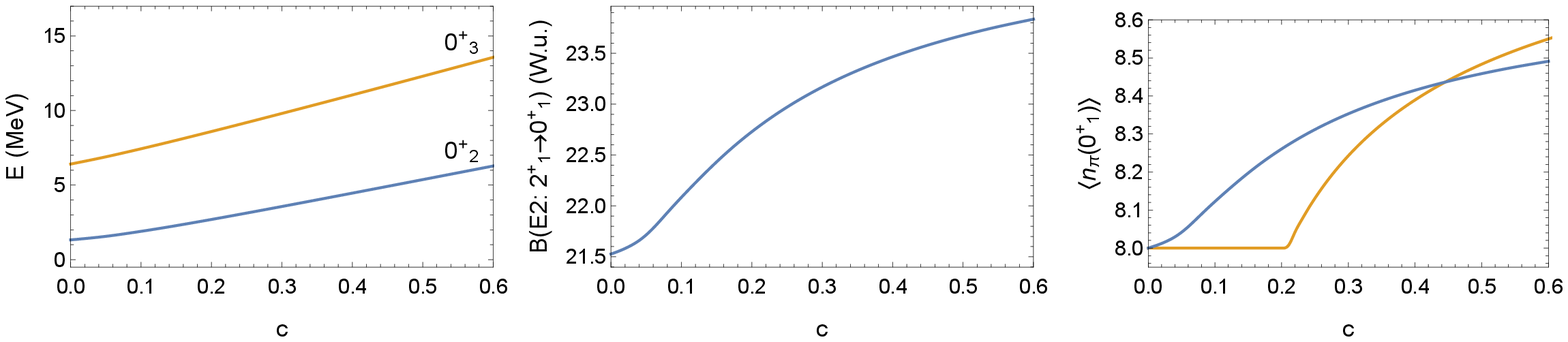}
\end{center}
\caption{Plots of various quantities for the system ${}^{16}\mathrm{O}+\alpha\to{}^{20}\mathrm{Ne}$ as functions of parameter $c$. In the left we plot the first $0^+$ states. In the middle we plot the transition probability $B(E2)$ of the state $2^{+}_1$ to the state $0^{+}_1$. In the right we plot the expectation value of the number of ${\bd n}_{\pi}$ bosons in the ground state: obtained with the numerical calculation depicted as a blue line, and with the coherent states depicted as a yellow line. The phase transition occurs at about $c=0.2$, as seen in the right plot.}
\label{figure5}
\end{figure}

In the left plot of Fig. \ref{figure5} we can see that the change in the $0^{+}_2$ and $0^{+}_3$ states is now smooth and no avoided level crossings are found as the parameter $c$ increases. However, traits of the second order QPT can still be seen elsewhere. In the middle plot of Fig. \ref{figure5} we show the transition probabilities of the state $2^{+}_1$ to the ground state $0^{+}_1$ as a function of the parameter $c$, where we can see a noticeable change near a critical value of $c$. In the right plot of Fig. \ref{figure5} we show the expectation value of the number of $n_{\pi}$ bosons in the ground state $\langle {\bd n}_{\pi}(0^{+}_1) \rangle$ as a function of the parameter $c$, obtained with the numerical calculation (blue) and with the coherent states (yellow). A sharp change at about $c=0.2$, the point of the phase transition, is present in the coherent state plot, whereas the change in the numerical calculation is smoother. The classical treatment, thus, exhibits
more clearly the event of a phase transition than the
exact Quantum Mechanical treatment, which includes
continuous changes in the mixing of states between
the two minima of the potential. This example also shows
that the signature of a phase transition not always
shows up clearly in the spectrum and transition values. 

The behaviour seen is typical for finite quantum systems:
In a strict sense, finite systems cannot exhibit a phase
transition, however, a structural change
can happen, with an order parameter changing significantly
in a short range within the parameter space. 
In this example,
the continuous change is explained by the formation of the two minima, i.e., the wave-function of the ground state changes
its dominant contribution from 
the spherical minimum ($\alpha=0$) to the deformed minimum ($\alpha \neq 0$) in a continuous
way and no discrete jump is produced. This is different in 
the semi-classical description. There, a discrete jump is 
generated, the moment the global minimum at $\alpha =0$
changes to the deformed minimum at $\alpha \neq 0$.
Therefore, the
phase transition is more clearly seen using a 
semi-classical potential.

\section{Phase transitions in an effective model for
QCD at low energy}
\label{QCD}

The catastrophe theory not only can be applied to
a nuclear cluster model but also to a topic as different
as QCD. 
Though, QCD seems to be quite different to algebraic cluster
models,it is also a many-body problem and the knowledge, 
acquired before, can be directly extended to this new field.

The main problem in QCD is to describe the structure of
the vacuum, the lowest state in energy, which is not 
trivial at all.
Without an interaction, the vacuum structure is
simple, i.e., there are no quarks nor gluons present.
However, when the interaction is switched on, the
physical vacuum should contain a structure involving
quarks, antiquarks and gluons (see \cite{PR1985}).

\begin{figure}[ht]
\begin{center}
\includegraphics[scale=1.0]{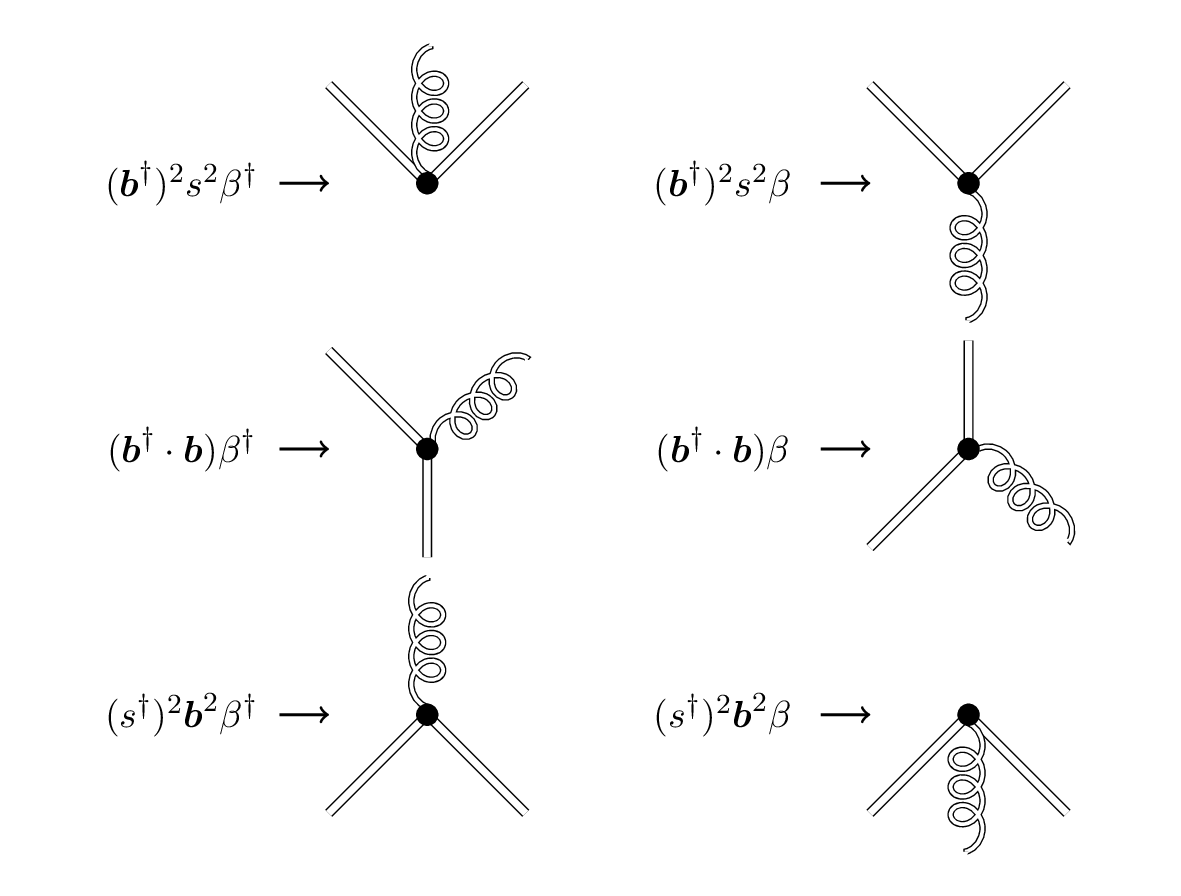}
\end{center}
\caption{Graphical diagrams for the interaction between
quark-antiquarks and gluon pairs, as they appear in the
Hamiltonian. They are motivated by the real QCD
interaction. To the left of each graph the corresponding
interaction is listed. A quark-antiquark is represented
by a double straight 
line and a gluon pair by a double-wavy line.
The time goes from bottom to top. 
}
\label{figure-int}
\end{figure}

Instead of using the real QCD,
effective models for QCD are easier to apply than a full
scale non-perturbative treatment, though these effective 
models are non-perturbative too. One such model was
proposed in \cite{QCD-0}, using a simple Hamiltonian
with a structure similar to QCD. In \cite{QCD-1,QCD-2}
a more sophisticated model was proposed, which we 
will use here in a modified form. 
The basic ingredients of this model
are {\it pairs} of quark-antiquarks and of 
gluon {\it pairs}. As seen in Fig. \ref{figure-int}
the interaction between those pairs are very similar
as in QCD between the quarks and gluons.

The model is able to describe the low lying meson spectrum
\cite{QCD-1}. In \cite{QCD-2} the evolution of the
{\it Quark Gluon Plasma} was described. It is interesting to
note that even particle-anti-particle production rates 
could be reproduced. In \cite{QCD-3} hadron states were
also described well, while also Penta-Quark and 
Hepta-Quark states were predicted.

\begin{figure}[ht]
\begin{center}
\includegraphics[scale=1.0]{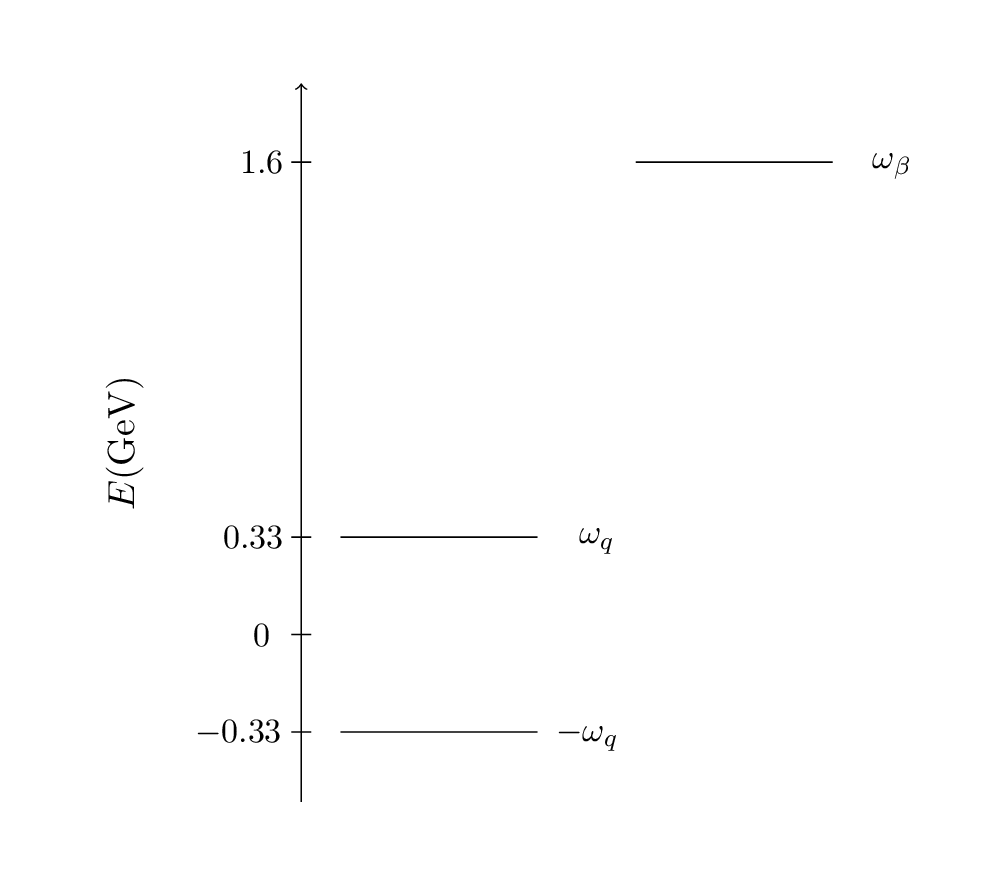}
\end{center}
\caption{This figure depicts the relative position
of the quark and gluon states. The quark sector is described
by a Lipkin model, consisting of 
two levels at $\pm \omega_q$.
The value of $\omega_q$ is $0.33\,\mathrm{GeV}$, i.e., three times of
that value reproduces approximately the mass of a nucleon. The
gluon state is at $1.6\,\mathrm{GeV}$, corresponding to 
the energy of two gluons, with the one-gluon energy
of $0.8\,\mathrm{GeV}$.
}
\label{energies}
\end{figure}

We follow the model described in 
\cite{QCD-1}, with a different trial state, implying
a modification in the structure of the interaction.
The energy scales of the model are depicted in Fig. 
\ref{energies},
which defines the scale of the fermion and boson states.
The scale of the fermionic state is 
$\omega_q=0.33\,\mathrm{GeV}$, a value chosen because three times this
value should give approximately the mass 
of the nucleon state, i.e.,
of $1\, \mathrm{GeV}$. 
Each fermionic state has a degeneracy of $2\Omega = 18$ (3 color, 3 flavour and 2 spin degrees of freedom).
A Dirac picture is employed: 
In the perturbative vacuum 
the lowest fermion level, at $-\omega_q$, is fully
occupied. Excited states are obtained by 
lifting quarks from 
the lower to the higher level, creating a particle-hole
state. A hole is described as an antiquark, i.e., a 
particle-hole state corresponds to a quark-antiquark
state.

The Hamiltonian of the model is 
\begin{eqnarray}\label{3.1}
{\bd H}&=&2\omega_q {\bd n}_q +\omega_{\beta} {\bd n}_{\beta} +C \left\{\left[({\bd b}^{\dagger})^2 s^2+2({\bd b}^{\dagger}\cdot {\bd b})+(s^{\dagger})^2 {\bd b}^2 \right]\left(1-\frac{{\bd n}_q}{2\Omega} \right)\beta^{\dagger}\right. \cr
&& \left. +\beta \left(1-\frac{{\bd n}_q}{2\Omega} \right)\left[({\bd b}^{\dagger})^2 s^2 +2({\bd b}^{\dagger}\cdot {\bd b})+(s^{\dagger})^2 {\bd b}^2  \right]  \right\}
~~~,
\label{ham}
\end{eqnarray}
where $n_q$ is the number of quark-antiquarks 
(particle-hole) {\it pairs}, 
$n_{\beta}$ the number gluon {\it pairs},
${\bd b}^\dagger$, ${\bd b}$ are the quark-antiquark
pair creation and annihilation operators and
${\bd \beta}^\dagger$, ${\bd \beta}$ are the same for 
the gluon pairs. The parameter $C$ gives the intensity
of the interaction. When $C=0$ there is no interaction
and the ground state is always given by all pairs of 
quarks in the lower level and no gluon pairs.
The factor $\left(1-\frac{{\bd n}_q}{2\Omega}\right)$
considers the PEP by shutting off the interaction when
all quarks are excited to the upper fermion level, thus,
no further excitation can take place. In Fig. \ref{figure-int} we depict the graphical presentation
of the interactions in (\ref{ham}) in terms of
double straight lines for the quark-antiquark pairs
and by double wavy lines for the gluon pairs.

The coherent state is defined as:
\begin{equation}\label{3.2}
|\psi \rangle = \mathcal{N}_{q}\mathcal{N}_{\beta} e^{\gamma \beta^{\dagger}}\left[s^{\dagger}+({\bd \alpha} \cdot {\bd b}^{\dagger}) \right]^{2\Omega}|0\rangle
~~~,
\end{equation}
where $\mathcal{N}_{q}$ and $\mathcal{N}_{\beta}$ 
are normalization factor for the fermions and bosons, respectively, given by
\beqa
{\cal N}_q & = & \frac{1}
{\left[(2 \Omega)!\left(1+\left({\bd \alpha}^*\cdot 
{\bd \alpha}\right)
\right)^{2 \Omega}
\right]^{\frac{1}{2}}}
\nonumber \\
{\cal N}_\beta & = & e^{-\frac{\gamma^*\gamma}{2}}
~~~.
\label{norm}
\eeqa
The vacuum state $|0\rangle = |0\rangle_{q}|0\rangle_{\beta}$ is the direct product of the fermion and boson vacuum. 
In \cite{QCD-0} a different coherent state was
used. The one in (\ref{3.2}) is inspired by a collective 
model in nuclei: The basic elements are pairs of fermions
which behave as bosons. The Pauli exclusion principle is
taken into account by the maximal number of particle-hole
pairs which can be created, as explained above.
One introduces an {\it auxiliary} 
scalar boson ($s$) and requires that the sum of $s$ and 
$b$ bosons is constant, given by $2\Omega$. 
In the perturbative vacuum the state below 0 is fully
occupied, which in the effective model corresponds to
having 18 quarks in the lowest state (Dirac picture). 
Because the total number of
particles in the Dirac picture is $2\Omega$
= $n_s+n_b$ ($n_s$ is the number of $s$-bosons and $n_b$
is the number of $b$-bosons), the number of quark-antiquark
{\it pairs} vary from 0 to $2\Omega$. 
The model does not include orbital degrees
of freedom.
If more quark-antiquarks
pairs are needed, one has to include these orbital degrees
of freedom. 

In a similar fashion
we obtain the semi-classical potential $V=\langle {\bd H}\rangle$ as the expectation value of the Hamiltonian (\ref{3.1}) in the basis of coherent states (\ref{3.2}):
\begin{eqnarray}\label{3.3}
V&=& 2\omega_q (2\Omega)\frac{({\bd \alpha}^{*}\cdot {\bd \alpha})}{\left[1+({\bd \alpha}^{*}\cdot {\bd \alpha})\right]^2}+\omega_{\beta}\gamma^{*}\gamma + 2\Omega (2\Omega -1)C (\gamma + \gamma^{*}) \left[\frac{({\bd \alpha}\cdot {\bd \alpha})+({\bd \alpha}^{*}\cdot{\bd \alpha}^{*})}{\left[1+({\bd \alpha}^{*}\cdot {\bd \alpha})\right]^2} \right.\cr
&&\left. +\frac{2}{2\Omega -1} \frac{({\bd \alpha}^{*}\cdot {\bd \alpha})}{\left[1+({\bd \alpha}^{*}\cdot {\bd \alpha})\right]^2} \right] \left(1-\frac{({\bd \alpha}^{*}\cdot {\bd \alpha})}{1+({\bd \alpha}^{*}\cdot {\bd \alpha})}\right) 
~~~.
\end{eqnarray}

Using the following parametrization of the coherent states:
\begin{eqnarray}\label{3.4}
({\bd \alpha}\cdot{\bd \alpha}) &=&  \alpha_{(0,0)}^2 e^{2i\phi_{(0,0)}}+\alpha_{(1,1)}^2 e^{2i\phi_{(1,1)}}\cr
({\bd \alpha}^{*}\cdot{\bd \alpha}) &=& \alpha_{(0,0)}^2 + \alpha_{(1,1)}^2\cr
({\bd \alpha}^{*}\cdot{\bd \alpha}^{*}) &=& (\alpha_{(0,0)}^{*})^2 e^{-2i\phi_{(0,0)}} + (\alpha_{(1,1)}^{*})^2 e^{-2i\phi_{(1,1)}} \cr
\gamma &=& \gamma e^{i\phi_{\gamma}} 
\end{eqnarray}
we can rewrite the semi-classical potential as
\begin{eqnarray}\label{3.5}
 V&=& 2\omega_q (2\Omega)\frac{\alpha_{(0,0)}^2+\alpha_{(1,1)}^2}{\left[1+\alpha_{(0,0)}^2+\alpha_{(1,1)}^2\right]^2} +\omega_{\beta}\gamma^{2}\cr
&&+ 2\Omega (2\Omega -1)C \gamma (e^{i\phi_{\gamma}}+e^{-i\phi_{\gamma}}) \left[\frac{\alpha_{(0,0)}^2(e^{2i\phi_{(0,0)}}+e^{-2i\phi_{(0,0)}})+\alpha_{(1,1)}^2(e^{2i\phi_{(1,1)}}+e^{-2i\phi_{(1,1)}})}{\left[1+\alpha_{(0,0)}^2+\alpha_{(1,1)}^2\right]^2} \right.\cr
&&\left. +\frac{2}{2\Omega -1} \frac{(\alpha_{(0,0)}^2+\alpha_{(1,1)}^2}{1+\alpha_{(0,0)}^2+\alpha_{(1,1)}^2} \right]\left(1-\frac{\alpha_{(0,0)}^2+\alpha_{(1,1)}^2}{1+\alpha_{(0,0)}^2+\alpha_{(1,1)}^2}\right) \cr
 V &=& 2 \omega_q (2\Omega) \frac{\alpha^2}{(1+\alpha^2)^2}+\omega_{\beta}\gamma^2 \cr
 &&+2\Omega(2\Omega -1)C 2\gamma \cos\phi_{\gamma} \left[\frac{2 \alpha_{(0,0)}^2 \cos 2\phi_{(1,1)} + 2\alpha_{(1,1)}^2\cos 2\phi_{(1,1)} }{(1+\alpha^2)^2}\right.\\
&&\left. +\frac{2}{2\Omega -1} \frac{\alpha^2}{1+\alpha^2} \right] \left(1-\frac{\alpha^2}{1+\alpha^2} \right),\nonumber
\end{eqnarray}
where we defined $\alpha_{(0,0)}^2+\alpha_{(1,1)}^2\equiv
\alpha^2$ as a new variable. The $\alpha_{(\lambda , \mu )}$
refer to the coupling of an quark-antiquark to a definite
flavour irrep, where $(0,0)$ is flavour zero and $(1,1)$ is the
flavour octet. The structure of the potential is similar to the
one of the SACM, save with simpler 
$F_{pq}(\alpha )$-functions. For a given $\alpha^2$, the
$\alpha_{(0,0)}^2$ can vary from 0 to $\alpha^2$ and so
in the opposite manner also the $\alpha_{(1,1)}^2$, which
is a symmetry property of the model.

We now continue with the minimization of the potential. The critical points of the angular variables are easily obtained, i.e., they are given by 
$\phi_{\gamma}=\phi_{(0,0)}=\phi_{(1,1)}=0$. The semi-classical potential is then a function of two variables:
\begin{equation}\label{3.6}
 V(\alpha,\gamma; C) = 4\Omega \omega_q  \frac{\alpha^2}{(1+\alpha^2)^2}+\omega_{\beta}\gamma^2 +8\Omega(2\Omega -1)C \gamma \frac{\alpha^2}{(1+\alpha^2)^2} \left[\frac{1}{1+\alpha^2}+\frac{2}{2\Omega -1}  \right].
\end{equation}

The potential has a quadratic dependence on the variable $\gamma$. The critical points of the potential satisfy $\nabla V(\alpha_c,\gamma_c)=0$. The component for the variable $\gamma$ is the partial derivative
\begin{equation}\label{3.7}
\left.\frac{\partial V}{\partial \gamma}\right|_{\gamma_c} = 2\omega_{\beta} \gamma_c + 8\Omega(2\Omega-1)C\frac{\alpha^2}{(1+\alpha^2)^2}\left[\frac{1}{1+\alpha^2}+\frac{2}{2\Omega -1}  \right] =0 ,
\end{equation}
and solving for $\gamma_c$ we obtain
\begin{equation}\label{3.8}
\gamma_c = -4\Omega (2\Omega -1)\frac{C}{\omega_{\beta}}\frac{\alpha^2}{(1+\alpha^2)^2}\left[\frac{1}{1+\alpha^2}+\frac{2}{2\Omega -1}  \right].
\end{equation}
Direct substitution of (\ref{3.8}) in (\ref{3.6}) permits us to write the one dimensional potential as
\begin{equation}\label{3.9}
V(\alpha,\gamma_c;C) = 4\Omega \omega_q \left\{\frac{\alpha^2}{(1+\alpha^2)^2} - \kappa\frac{\alpha^4}{(1+\alpha^2)^4}\left[\frac{1}{1+\alpha^2}+\frac{2}{2\Omega -1}  \right]^2 \right\} ~~~.
\end{equation}
The dimensionless parameter $\kappa$ is defined as
\begin{equation}\label{3.10}
\kappa = 4\Omega (2\Omega -1)^2\frac{C^2}{\omega_q\omega_{\beta}}.
\end{equation}

\begin{figure}[ht]
\begin{center}
\includegraphics[scale=0.8]{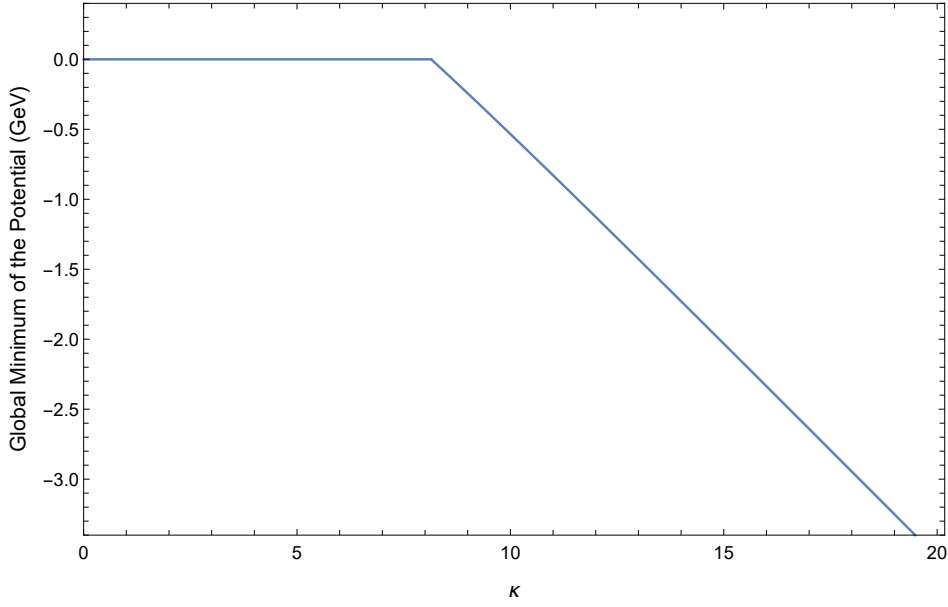}
\end{center}
\caption{Global minimum of the semi-classical potential as a function of $\kappa$. At $\kappa=8.14502$ ($C=0.020331 \,\mathrm{GeV}$) a first order phase transition occurs. A discontinuity in the first derivative with respect to $\kappa$ can be seen in the plot. We used the values $\omega_q =0.33 \,\mathrm{GeV}$, $\omega_{\beta}=1.6 \,\mathrm{GeV}$ and $2\Omega = 18$.}
\label{figure8}
\end{figure}

\subsection{Bifurcation and Maxwell sets}
In the one dimensional parameter space $\kappa$ the bifurcation and Maxwell sets can be obtained by direct application of the formulas in Appendix A.

The critical manifold is the surface of critical points spanned by the variation of parameter $\kappa$. The critical points are those that satisfy
\begin{equation}\label{3.11}
\frac{\partial V (\alpha,\gamma_c;C)}{\partial \alpha} = 0.
\end{equation}

This potential exhibits an interesting feature. The Hessian determinant of the potential (\ref{3.6}) at $\alpha=0$ and $\gamma =0$ is given by
\begin{eqnarray}
\left| \begin{array}{cc}
\frac{\partial^2 V (\alpha,\gamma;C)}{\partial \alpha^2} & \frac{\partial^2 V (\alpha,\gamma;C)}{\partial \alpha \partial \gamma} \\
\frac{\partial^2 V (\alpha,\gamma;C)}{\partial\gamma \partial \alpha} & \frac{\partial^2 V (\alpha,\gamma;C)}{\partial \gamma^2}
\end{array} \right|_{(\alpha=0,\gamma=0)} = 16\Omega \omega_q\omega_{\beta}
~~~,
\end{eqnarray}
and we can see that it is always positive. In other words, the semi-classical treatment gives a potential which maintains a local minimum at the origin $(\alpha =0,\gamma =0)$, even after the phase transitions, which corresponds to the perturbative vacuum. A local minimum is related to an excited state, i.e. the perturbative vacuum still persists as an excited state at higher energies. 

We solve (\ref{3.11}) for $\kappa$ and get
\begin{equation}\label{3.12}
\kappa = - \frac{(1-\alpha^2)(1+\alpha^2)^4(2\Omega-1)^2}{2\alpha^2(\alpha^2+2\Omega)(\alpha^4+2\alpha^2(2\Omega-1)-2\Omega)}.
\end{equation} 
The singular mapping to the parameter space happens when the derivative of (\ref{3.12}) with respect to $\alpha$ is zero. This condition leads to a 10th degree polynomial:
\begin{eqnarray}\label{3.13}
f_B(\alpha,2\Omega) &=& -4\Omega^2+ 4\Omega \alpha^2 (7\Omega -3)-2\alpha^4 (20\Omega^2 -18\Omega +3)+2\alpha^6 (12\Omega^2 -18\Omega +5)\cr
&&+\alpha^8(12\Omega -7)+\alpha^{10}
~~~,
\end{eqnarray}
whose solution is to be substituted in (\ref{3.12}) to get 
the bifurcation set. For the set values of $\omega_q =0.33 \,\mathrm{GeV}$, $\omega_{\beta}=1.6 \,\mathrm{GeV}$ and $2\Omega = 18$, we obtain
\begin{equation}\label{3.14}
\kappa_B=5.8997 ,
\end{equation}
and using the definition (\ref{3.10}) we return to the original parameter $C$:
\begin{equation}\label{3.15}
C_{B}=0.017303 \,\mathrm{GeV} .
\end{equation} 

To obtain the Maxwell set we consider the roots of the potential
\begin{equation}\label{3.16}
V(\alpha, \gamma_c;C)+V_0 = 0,
\end{equation}
and define the roots manifold as the surface spanned by the 
variation of parameter $\kappa$ of all the real roots 
satisfying (\ref{3.16}). The Maxwell set will be the one
obtained where the mapping of this surface to the parameter 
space is singular.

We solve for $\kappa$ in (\ref{3.16}), setting $V_0=0$ (as no two 
minima coincide at $V_0\neq 0$ for the potential (\ref{3.9})), and get
\begin{equation}\label{3.17}
\kappa =  \frac{(1+\alpha^2)^4(2\Omega-1)^2}{\alpha^2(\alpha^2+2\Omega)^2}.
\end{equation} 
The singular mapping to the parameter space happens when the derivative of (\ref{3.17}) with respect to $\alpha$ is zero. This condition leads to a 4th degree polynomial:
\begin{equation}\label{3.18}
f_M(\alpha,2\Omega) = -2\Omega+ 3\alpha^2 (2\Omega-1)+\alpha^4
~~~.
\end{equation}
The appropriate real solutions are given by
\begin{equation}\label{3.19}
\alpha=
\pm\sqrt{\frac{1}{2}\left(-3(2\Omega -1)+\sqrt{36\Omega^2-28 \Omega+9}\right)} 
\end{equation}
and substituting in (\ref{3.17}) we 
obtain the Maxwell set as a function of $2\Omega$. For completeness we write the explicit expression:
\begin{equation}\label{3.20}
\kappa_M(2\Omega) = \frac{(2\Omega-1)^2\left(5-6\Omega+\sqrt{9+4\Omega(9\Omega-7)}\right)^4}{2\left(3-6\Omega+\sqrt{9+4\Omega(9\Omega-7)}\right)\left(3-2\Omega+\sqrt{9+4\Omega(9\Omega-7)}\right)^2}
~~~.
\end{equation}
For the set values of $\omega_q =0.33 \,\mathrm{GeV}$, $\omega_{\beta}=1.6 \,\mathrm{GeV}$ and $2\Omega = 18$, we get
\begin{equation}\label{3.21}
\kappa_M=8.14502 ,
\end{equation}
and using the definition (\ref{3.10}) we return to the original parameter $C$:
\begin{equation}\label{3.22}
C_{M}=0.020331 \,\mathrm{GeV} .
\end{equation} 

In Fig. \ref{figure8} we plot the global minimum of the potential as a function of $\kappa$. The point of the phase
transition is given by the kink a little above $\kappa=8$ ($C=0.02\,\mathrm{GeV}$).
At this point, a former excited $0^{++}$
(The notation is $J^{PC}$, with the spin $J$,
parity $P$ and charge conjugation $C$) state, 
with the same
quantum numbers as the perturbative vacuum, crosses
the perturbative vacuum, being the new vacuum state, which contains a finite number of quark-antiquark pairs and pairs of gluons.

\begin{figure}[ht]
\begin{center}
\includegraphics[scale=0.8]{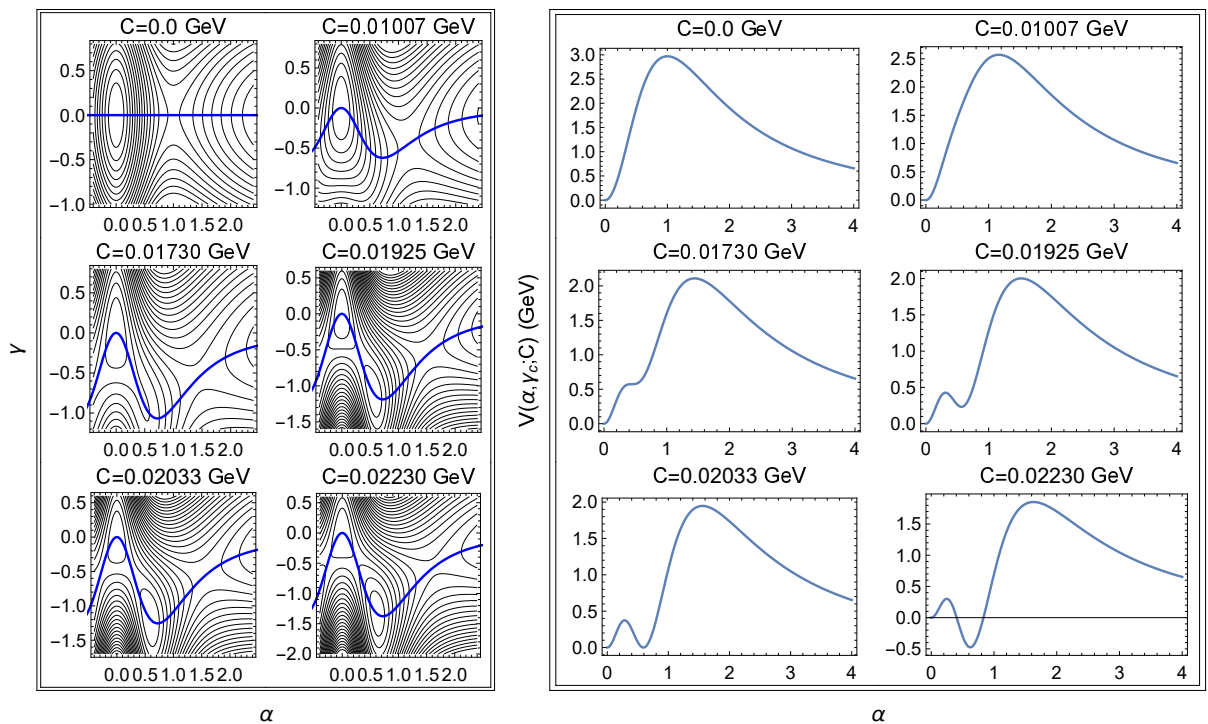}
\end{center}
\caption{Contour plots $(\alpha,\gamma)$ of the semi-classical potential (\ref{3.6}) for different values of parameter $C$ with their corresponding one dimensional semi-classical potential (\ref{3.9}). The blue line in the left panel corresponds to the critical values of $\gamma_c$ as a function of $\alpha$ given in (\ref{3.8}). We used the values $\omega_q =0.33 \,\mathrm{GeV}$, $\omega_{\beta}=1.6 \,\mathrm{GeV}$ and $2\Omega = 18$.}
\label{figure9}
\end{figure}

In Fig. \ref{figure9} we show the contour plots of the two-dimensional semi-classical potential (\ref{3.6}) for different values of $C$, along with the critical value for $\gamma$ given in (\ref{3.8})
and indicated by a solid blue line. 
The corresponding one-dimensional semi-classical 
potentials (\ref{3.9}) are also plotted. As the intensity of the parameter increases, passing the critical value $C_M$ (\ref{3.22}), the global minimum is no longer at $\alpha=0$ and a first order quantum phase transition occurs. The jump occurs at 
the same point as in Fig. \ref{figure8}. 

For small $C$ the global minimum is at $\alpha = 0$, where we
do not take into account the asymptotic minimum for
$\alpha \rightarrow \infty$. Then, with increasing $C$
a second minimum forms, first at larger energies than the one
at $\alpha = 0$. The Maxwell point is reached when both 
minima are at the same height. For even larger $C$
the global minimum is a deformed one ($\alpha > 0$), but
still maintaining an excited minimum at $\alpha = 0$.
A local minimum indicates the existence of the state, i.e.,
the model predicts that after the phase transition the
former perturbative vacuum state still exists as an excited
state. This is in accordance to \cite{QCD-1}, where the
$f_0(400-1200)$ state has a content of half a quark-antiquark
and half a gluon pair (the non-integer number refers to the
expectation number of the pair operators with respect to
the state function).
That such a property emerges in this particular model
is a novelty, indicating that the use of coherent states
can give more information than just the ground state
properties. 

\subsubsection{Some consequences}

\begin{figure}[ht]
\begin{center}
\includegraphics[scale=0.8]{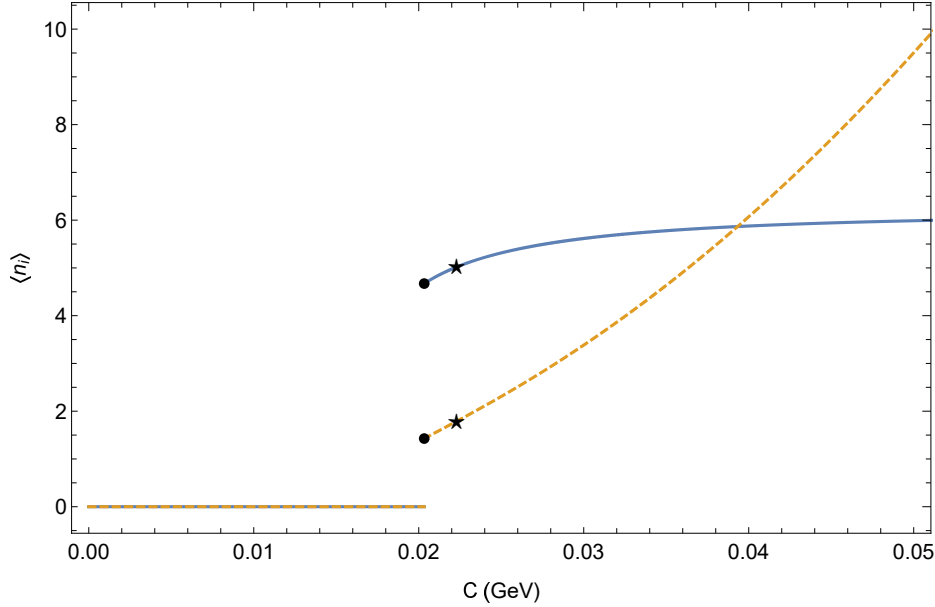}
\end{center}
\caption{Expectation values of quarks pairs $\langle {\bd n}_q\rangle$ (blue, solid) and gluon pairs $\langle {\bd n}_{\beta}\rangle$ (yellow, dashed) as a function of the parameter $C$. The dots represent the number of quarks pairs and gluon pairs at the Maxwell set $C=0.020331 \,\mathrm{GeV}$. The stars represent the number of quarks pairs and gluon pairs at $C=0.022301 \,\mathrm{GeV}$. We used the values $\omega_q =0.33 \,\mathrm{GeV}$, $\omega_{\beta}=1.6 \,\mathrm{GeV}$ and $2\Omega = 18$.}
\label{figure10}
\end{figure}

The expectation number of the quarks and gluons pairs are, respectively:
\begin{eqnarray}\label{0.1}
\langle {\bd n}_q\rangle &=& 2\Omega \frac{\alpha^2}{1+\alpha^2} \cr
\langle {\bd n}_{\beta}\rangle &=& \gamma^2 .
\end{eqnarray}
In Fig. \ref{figure10} these values are plotted as a function of parameter $C$. The plots are obtained by evaluating $\langle {\bd n}_q\rangle$ and $\langle {\bd n}_{\beta}\rangle$ at the critical point $(\alpha_c, \gamma_c)$ of the global minimum of the semi-classical potential. A discontinuity is present at the point of the first order phase transition 
$C=0.020331 \,\mathrm{GeV}$. At about $C=0.039 \,\mathrm{GeV}$ the gluon number surpasses the quarks.
With increasing interaction ($C$), the number of 
quark-antiquark pairs becomes saturated, while the gluon 
pairs continue to rise, winning over the 
quark-antiquark pairs.

In \cite{IJMPE2006} a relation of the quark-antiquark
and gluon condensate \cite{PR1985} to the number
of quark-antiquark and gluon pairs in the physical
vacuum $| {\rm vac}\rangle$ was given, namely
\beqa
\langle {\rm vac} | {\bar \Psi}_f \Psi_f | {\rm vac}
\rangle & = & \frac{1}{V} \left( \frac{2n_q}{3}-6\right)
\nonumber \\
\langle {\rm vac} | \frac{\alpha_s}{\pi}
F_{\mu\nu}^a F^{\mu\nu}_a | {\rm vac}
\rangle & = & \left( \frac{\alpha_s 16 \pi}
{\omega_\beta^2 V^2}
\right) \left( 4 n_\beta + 9\right)
~~~.
\label{condensate}
\eeqa
The $\Psi_f$ is the fermion function and 
$V=\frac{4\pi}{3}r_0^3$ is the volume of the 
size of a hadron with radius $r_0=0.875\,\mathrm{fm}$ =
$4.375\,\mathrm{GeV}^{-1}$
\cite{IJMPE2006}.
The $\alpha_s$ is the strong coupling constant.
The values of the quark and gluon condensates in
\cite{PR1985} are
\beqa
\langle {\rm vac} | {\bar \Psi}_f \Psi_f | {\rm vac}
\rangle & = & -(0.223\,{\rm GeV})^3
\nonumber \\
\langle {\rm vac} | \frac{\alpha_s}{\pi}
F_{\mu\nu}^a F^{\mu\nu}_a | {\rm vac}
\rangle & = & (0.360\,{\rm GeV})^4
~~~.
\label{val-cond}
\eeqa
The first condition with (\ref{condensate}) leads to
$n_q\approx 4$. Comparing this number to Fig.
\ref{figure10} we can identify this number to be 
approximately
realized just after the phase transition took place. 
Thus we selected two points near this value: 
In Fig. \ref{figure10} two positions just
after the phase transition are indicated by a dot
(with $C_M=C_1=0.020331 \,\mathrm{GeV}$) and
by a star (with $C_2=0.022301 \,\mathrm{GeV}$). 
For the point $C_M$ the expectation values are
$\langle n_q\rangle =4.67192$, 
$\langle n_\beta \rangle = 1.42697$. 
They lead to $-(0.202\,{\rm GeV})^3$ for
the quark condensate. For the gluon condensate
we solve for $\alpha_s$, obtaining $\alpha_s = 7.156$,
not very far from the value deduces in 
\cite{IJMPE2006} and also in good agreement with 
\cite{PR1985}. For the star value $C_2$ the same procedure,
with $\langle n_q\rangle =5.01946$ and 
$\langle n_\beta \rangle = 1.79012$, 
leads to the quark condensate value $-(0.1963\,{\rm GeV})^3$,
a little lower than in the former case, and to the
strong coupling constant $\alpha_s = 6.513$, a little
lower than in the former case.

This last calculation shows that the model is consistent,
that the catastrophe theory is a helpful guide 
to describe the phase
transition and that the physical vacuum is 
probably a state near the
point of a phase transition.

\section{Conclusions}
\label{conclusions}

One motivation of this contribution was to extend a cluster model (the SACM) to include deformation dependencies in the clusters and to correlate signatures in the spectrum with phase transitions. Another motivation was to show that the catastrophe theory has a wide range of practical applications to different areas in physics, such as QCD, which was the second exampled discussed in this contribution. Applications to Optics and General Relativity were mention earlier in the Introduction.

In the case of the SACM we showed the effectiveness of the catastrophe theory to describe a very complicated semi-classical potential, where deformation effects were included. 
Phase transitions up to 3rd order were encountered,
depending also on the path taken in the phase space.
It was possible to identify quantum phase transitions effects when a numerical calculation of the spectrum was performed. Changes in the spectrum as a function of an interaction parameter corresponded to a transition from one region in parameter space to another, where the global minimum of the semi-classical potential changes from spherical to deformed. It was found 
that the changes in the numerical calculations of the transition probability and the expectation value of the number of $n_{\pi}$ bosons were smooth, while in the semi-classical description the changes were clearly marked by a jump in the structure. This is due to the fact that in a Quantum Mechanical treatment the wave function changes smoothly from one minimum to the other, while in the semi-classical treatment the jump happens when one minimum is below the other.

As a second example the phase transition of an effective model of QCD was investigated, describing how the perturbative vacuum changes to a vacuum with a content of 
quark-antiquark pairs and gluons pairs. The phase transition
is of first order.

We hope that the motivation of the importance and effectiveness of the catastrophe theory formalism to the study of phase transitions, along with the techniques described here, will encourage the reader to apply it in other areas of interest.

\vskip 0.5cm
\section*{Acknowledgment}

P.O.H. acknowledges financial support from 
DGAPA-PAPIIT (IN100421). D.S.L.R. 
acknowledges financial support from a scholarship
(No. 728381) 
received from CONACyT.
E.L.M. acknowledges financial support from
DGAPA-PAPIIT (IN114821). 

\begin{appendix}
\section{General expressions for the bifurcation set and Maxwell set}\label{appendix1}
In some steps the software MATHEMATICA \cite{mat11}
was used to simplify some expressions.

Let us consider a one dimensional real function $V(x;r_1,r_2,r_3)$ dependent on three real parameters $r_i$ of the form
\begin{equation}\label{a.1}
V(x;r_1,r_2,r_3)=r_1g_1(x)+r_2g_2(x)+r_3g_3(x)+g_4(x),
\end{equation}
where $g_j(x)$, $j=1,2,3,4$, are arbitrary rational functions with no singularities in the domain of $x$.

The bifurcation set is the subspace in parameter space 
$(r_1,r_2,r_3)$, where the mapping $(x_c,r_1,r_2,r_3)\mapsto (r_1,r_2,r_3)$ of the critical manifold to the parameter space is singular, i.e., when the Jacobian determinant of this transformation vanishes \cite{gilmore}. The bifurcation set serves as the separatrix where critical points begin to emerge.

The critical manifold is the surface of all critical points spanned by the variation of parameters satisfying 
\begin{equation} \label{a.2}
\left.\frac{\dd V}{\dd x}\right|_{x=x_c}= r_1g'_1(x_c)+r_2g'_2(x_c)+r_3g'_3(x_c)+g'_4(x_c)
=0.
\end{equation}
From (\ref{a.2}) we can solve for $r_1$ and obtain
\begin{equation}\label{a.3}
r_1 = - \frac{1}{g'_1(x_c)}\left( r_2g'_2(x_c) + r_3g'_3(x_c)+g'_4(x_c)\right),
\end{equation}
which allows us to express the parameter $r_1$ as a function of the critical points $x_c$ and the parameters $r_2$ and $r_3$; thus, the mapping of the critical manifold to the parameter space becomes $(x_c,r_2,r_3)\mapsto (r_1,r_2,r_3)$. Then, the Jacobian determinant of the mapping of the critical manifold to the parameter space is given by
\begin{eqnarray}\label{a.3a}
\left| \begin{array}{ccc}
\frac{\partial r_1}{\partial x_c} & \frac{\partial r_1}{\partial r_2} & \frac{\partial r_1}{\partial r_3} \\
\frac{\partial r_2}{\partial x_c} & \frac{\partial r_2}{\partial r_2} & \frac{\partial r_2}{\partial r_3} \\
\frac{\partial r_3}{\partial x_c} & \frac{\partial r_3}{\partial r_2} & \frac{\partial r_3}{\partial r_3}
\end{array} \right| 
= \left| \begin{array}{ccc}
\frac{\partial r_1}{\partial x_c} & \frac{\partial r_1}{\partial r_2} & \frac{\partial r_1}{\partial r_3} \\
0 & 1 & 0 \\
0 & 0 & 1 \end{array} \right| = \frac{\partial r_1}{\partial x_c} .
\end{eqnarray}
The mapping is singular when the Jacobian determinant vanishes. Using (\ref{a.3}) we have
\begin{equation}\label{a.4}
\frac{\partial r_1}{\partial x_c} = \frac{1}{\left(g'_1(x_c) \right)^2} \Big(r_2 W(g'_2,g'_1)+r_3W(g'_3,g'_1)+W(g'_4,g'_1)  \Big) = 0,
\end{equation}
where $W(f,g)=f(x)g'(x)-f'(x)g(x)$ is the Wronskian determinant. We solve for $r_2$ in (\ref{a.4}) and obtain
\begin{equation}\label{a.5}
r_2 =- \frac{1}{W(g'_1,g'_2)} \Big(r_3 W(g'_1,g'_3)+W(g'_1,g'_4) \Big),
\end{equation}
as a function of parameter $r_3$ and of the critical points $x_c$. Direct substitution of (\ref{a.5}) in (\ref{a.3}) gives us $r_1$ as a function of parameter $r_3$ and of the critical points $x_c$ as
\begin{equation}\label{a.6}
r_1 = \frac{1}{W(g'_1,g'_2)} \Big(r_3 W(g'_2,g'_3)+W(g'_2,g'_4) \Big) .
\end{equation}
The parametric surface in three-dimensional space defined by (\ref{a.5}) and (\ref{a.6}):
\begin{equation}\label{a.7}
\mathcal{C}_B = \left\{\big(r_1(x_c;r_3),r_2(x_c;r_3),r_3\big) \mid x_c,r_3\in \mathbb{R} \right\}
\end{equation}
is the bifurcation set of the potential function (\ref{a.1}).

The Maxwell set is the subspace in parameter space where for at least two critical points $x_1$ and $x_2$ the following condition holds
\begin{equation}\label{a.8}
V(x_1;r_1,r_2,r_3) = V(x_2;r_1,r_2,r_3) = -V_c,
\end{equation}
i.e. the value of the function at two critical points $x_1$ and $x_2$ is the same and equal to $-V_c$, with $V_c\in \mathbb{R}$.
 
This persuade us to consider the roots manifold, defined as the manifold of all the real roots $x_r$ spanned by the variation of parameters satisfying 
\begin{equation}\label{a.9}
V(x_r;r_1,r_2,r_3) + V_c=0.
\end{equation}
When the parameters are varied arbitrarily a critical point occurs when two real roots (or a conjugate complex pair) of (\ref{a.9}) coalesce. Thus, we consider the case when the mapping of the roots manifold (\ref{a.9}) to the parameter space is singular, which corresponds precisely to the coalescence of two roots. We solve for $r_1$ in (\ref{a.9}) and obtain:
\begin{equation}\label{a.10}
r_1= -\frac{1}{g_1(x_r)}\Big(r_2 g_2(x_r)+r_3g_3(x_r)+g_4(x_r)+V_c \Big).
\end{equation}
Analogous to (\ref{a.3a}), the Jacobian determinant of the transformation is 
\begin{equation}\label{a.11}
\frac{\partial r_1}{\partial x_r} = \frac{1}{g_1^2(x_r)}\Big(r_2 W(g_2,g_1) + r_3 W(g_3,g_1) + W(g_4,g_1) + g'_1(x_r)V_c \Big),
\end{equation}
and the singular mapping occurs when it vanishes. 
This allows us to solve for $r_2$ and obtain
\begin{equation}\label{a.12}
r_2 = - \frac{1}{W(g_1,g_2)}\Big(r_3 W(g_1,g_3) + W(g_1,g_4) -g'_1(x_r)V_c \Big),
\end{equation}
which is given in terms of the critical points $x_r$ and the parameter $r_3$ and $V_c$. Direct substitution of (\ref{a.12}) in (\ref{a.10}) allows us to obtain $r_1$ in terms of the same values as $r_2$:
\begin{equation}\label{a.13}
r_1 = \frac{1}{W(g_1,g_2)}\Big(r_3 W(g_2,g_3)+ W(g_2,g_4)-g'_2(x_r)V_c \Big).
\end{equation}
Equations (\ref{a.12}) and (\ref{a.13}) 
provide the values of $r_1$ and $r_2$ for which there exists a critical point $x_r$ such that $V(x_r;r_1,r_2,r_3)=-V_c$, for $r_3$ fixed. We can show that the $x_r$, which satisfy both (\ref{a.12}) and (\ref{a.13}), are critical points by direct substitution in (\ref{a.2}) with $x_c=x_r$.

Condition (\ref{a.8}) for the Maxwell set is equivalent to demand that there exist two different values $x_1$ and $x_2$ for which the following set of algebraic equations are simultaneously satisfied:
\begin{eqnarray}
r_2(x_1;r_3,V_c)&=&r_2(x_2;r_3,V_c)\label{a.14}\\
r_1(x_1;r_3,V_c)&=&r_1(x_2;r_3,V_c)\label{a.15},
\end{eqnarray}
with $r_2$ and $r_1$ given by (\ref{a.12}) and (\ref{a.13}), respectively. We can rewrite (\ref{a.14}) and (\ref{a.15}) as the following set of equations:
\begin{eqnarray}
r_2 &:& r_3 W_{13}(x_1,x_2)+ W_{14}(x_1,x_2)-V_c G_1(x_1,x_2)=0 \label{a.16}\\
r_1 &:& r_3 W_{23}(x_1,x_2)+ W_{24}(x_1,x_2)-V_c G_2(x_1,x_2)=0 \label{a.17}
\end{eqnarray}
where we defined the following functions:
\begin{eqnarray}
W_{ij}(x_1,x_2) &=& \left. W(g_i,g_j)\right|_{x_1} \left. W(g_1,g_2)\right|_{x_2} - \left. W(g_i,g_j)\right|_{x_2} \left. W(g_1,g_2)\right|_{x_1} \label{a.18}\\
G_{i}(x_1,x_2) &=& g'_i(x_1)\left. W(g_1,g_2)\right|_{x_2} -g'_i(x_2)\left. W(g_1,g_2)\right|_{x_1}\label{a.19} .
\end{eqnarray}

The system of equations (\ref{a.16}) and (\ref{a.17}) can be solved by eliminating $V_c$ and obtaining $r_3$ as a function of $x_1$ and $x_2$, and also by eliminating $r_3$ and obtaining $V_c$ as a function of $x_1$ and $x_2$. Eliminating $V_c$ from (\ref{a.16}) and (\ref{a.17}) we obtain:
\begin{eqnarray}\label{a.20}
&&r_3 \Big(G_2(x_1,x_2) W_{13}(x_1,x_2)-G_1(x_1,x_2) W_{23}(x_1,x_2) \Big) \cr
&&+ G_2(x_1,x_2) W_{14}(x_1,x_2)-G_1(x_1,x_2) W_{24}(x_1,x_2)  =0,
\end{eqnarray}
and eliminating $r_3$ from (\ref{a.16}) and (\ref{a.17}) we get:
\begin{eqnarray}\label{a.21}
&&V_c \Big(G_1(x_1,x_2)W_{23}(x_1,x_2)-G_2(x_1,x_2)W_{13}(x_1,x_2) \Big)\cr
&&+ W_{13}(x_1,x_2)W_{24}(x_1,x_2)-W_{23}(x_1,x_2)W_{14}(x_1,x_2)=0.
\end{eqnarray}
The combinations of functions appearing in (\ref{a.20}) and (\ref{a.21}) can be written as follows:
\begin{eqnarray}\label{a.22}
&&G_2(x_1,x_2) W_{1i}(x_1,x_2)-G_1(x_1,x_2) W_{2i}(x_1,x_2) \\
&&= \left. W(g_1,g_2)\right|_{x_1} \left. W(g_1,g_2)\right|_{x_2} \Big(g'_1(x_1)\left.W(g_2,g_i)\right|_{x_2}+g'_1(x_2)\left.W(g_2,g_i)\right|_{x_1}\cr
&&\quad -g'_2(x_1)\left.W(g_1,g_i)\right|_{x_2}-g'_2(x_2)\left.W(g_1,g_i)\right|_{x_1}+g'_i(x_1)\left.W(g_1,g_2)\right|_{x_2} +g'_i(x_2)\left.W(g_1,g_2)\right|_{x_1} \Big),\nonumber
\end{eqnarray}
and
\begin{eqnarray}\label{a.23}
&&W_{23}(x_1,x_2)W_{14}(x_1,x_2) -W_{13}(x_1,x_2)W_{24}(x_1,x_2)\cr
&&= -\left. W(g_1,g_2)\right|_{x_1} \left. W(g_1,g_2)\right|_{x_2}\Big(\left.W(g_1,g_2)\right|_{x_1}\left.W(g_3,g_4)\right|_{x_2}+ \left.W(g_1,g_2)\right|_{x_2}\left.W(g_3,g_4)\right|_{x_1}\cr
&&\quad -\left.W(g_1,g_3)\right|_{x_1}\left.W(g_2,g_4)\right|_{x_2}-\left.W(g_1,g_3)\right|_{x_2}\left.W(g_2,g_4)\right|_{x_1}\cr
&&\quad +\left.W(g_1,g_4)\right|_{x_1}\left.W(g_2,g_3)\right|_{x_2}+\left.W(g_1,g_4)\right|_{x_2}\left.W(g_2,g_3)\right|_{x_1}\Big). 
\end{eqnarray}
Note that (\ref{a.22}) and (\ref{a.23}) simplify both (\ref{a.20}) and (\ref{a.21}) because the common term $\left. W(g_1,g_2)\right|_{x_1} \left. W(g_1,g_2)\right|_{x_2}$ can be factored out of the equations.

Once $x_1$ and $x_2$ are determined from (\ref{a.20}), for a given value of $r_3$, we obtain the respective value of $V_c$ from (\ref{a.21}). Finally, we substitute it all back in (\ref{a.12}) and (\ref{a.13}), so that now $r_1$ and $r_2$ are given in terms of $x_1$, $x_2$ and $r_3$, and obtain the Maxwell set as the parametric surface in three-dimensional space defined as:
\begin{equation}\label{a.24}
\mathcal{C}_M = \left\{(r_1(x_1,x_2;r_3),r_2(x_1,x_2;r_3),r_3) \mid x_1,x_2,r_3 \in \mathbb{R} \right\},
\end{equation}
with $x_1$ and $x_2$ satisfying (\ref{a.14}) and (\ref{a.15}). In the simple case where the function (\ref{a.1}) has an extremum at $V_c=0$, it suffices to use (\ref{a.12}) and (\ref{a.13}).

\end{appendix}

\end{document}